\documentclass[preprint,showpacs,preprintnumbers,amsmath,amssymb]{revtex4}
\usepackage{graphicx}
\usepackage{graphics}
\usepackage{dcolumn}
\usepackage{bm}
\begin{document}

\title{Correlated photon-pair emission from \\
pumped-pulsed quantum dots embedded in a microcavity}

\author{J.I. Perea} \affiliation{Departamento de F\'{\i}sica
Te\'orica de la Materia Condensada, Universidad  Aut\'onoma de
Madrid, Cantoblanco 28049 Madrid, Spain.}

\author{F. Troiani}
\affiliation{Departamento de F\'{\i}sica Te\'orica de la Materia
Condensada, Universidad  Aut\'onoma de Madrid, Cantoblanco 28049
Madrid, Spain.}

\author{C. Tejedor}
\affiliation{Departamento de F\'{\i}sica Te\'orica de la Materia
Condensada, Universidad Aut\'onoma de Madrid, Cantoblanco 28049
Madrid, Spain.}

\begin{abstract}
We theoretically investigate the optical response of a quantum dot, embedded
in a microcavity and incoherently excited by pulsed pumping.
The exciton and biexciton transition are off-resonantly coupled with the
left- and right-polarized mode of the cavity, while the two-photon resonance
condition is fulfilled. Rich behaviours are shown to occur in the time
dependence of the second-order correlation functions which refer to
counter-polarized photons.
The corresponding time-averaged quantities, which are accessible to
experiments, confirm that such a dot-cavity system behaves as a good
emitter of single, polarization-correlated photon pairs.
\end{abstract}


\maketitle

\section{INTRODUCTION}
Semiconductor quantum dots (QDs) were recently demonstrated to
emit highly nonclassical light, under proper exciting
conditions~\cite{moreau,santori}. Besides being interesting in its
own right, such phenomena can be exploited for the implementation
of solid-state quantum-information devices~\cite{nielsen}. Among
other things, these require high collection efficiencies and
photon emission rates larger than the relevant decoherence ones:
two properties which are typically not present in individual QDs,
but are significantly approached when the QD is embedded in a
microcavity (MC). In fact, cavity photons are mostly emitted with
a high degree of directionality, thus allowing for large
collection efficiencies~\cite{solomon}. Moreover, the reduction of
the exciton recombination time (Purcell effect) minimizes the
decoherence due to exciton-phonon scattering
~\cite{gerard,solomon,unitt,varoutsis}. The experimental results
obtained in the low-excitation limit (i.e., at most one exciton at
a time in the QD) are relatively well
understood~\cite{moreau01,gerard,santori01,zwiller02,kiraz02,michler,pelton,fattal04}.
The QD optical selection rules and its discrete energy spectrum
can however be exploited also for emitting polarization-correlated
photon pairs. The main goal of the present work is that of
investigating the time-dependence and average values of the photon
correlations, in a condition where the pair-emission probability
is enhanced by suitably configuring the coupling to the MC.

The paper is organized as follows: in Sec.~II we describe the system and the
model we use; in Sec.~III, we present the results for the time-dependent
photon-coherence functions, as well as time-averaged quantities, which may
directly be compared with experimental results. Finally, we drow our
conclusions in Sec.~IV.

\section{ THE SYSTEM AND THE MODEL}

\begin{figure}
\includegraphics[clip,height=8cm,width=8.cm]{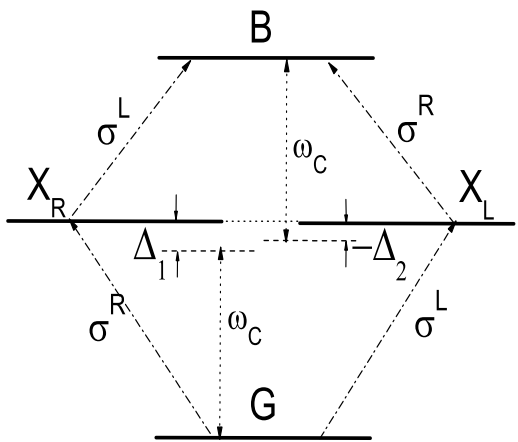}
\caption{Scheme showing the cavity-mode frequency and the QD
levels. The arrows show the two types of circularly polarized
optical transitions connecting the QD levels.} \label{lscheme}
\end{figure}

The discrete nature of the QD energy spectra significantly reduces
the portion of the dot Hilbert space which is required for the dot
description, at least in the low-excitation regime. Throughout the
paper, we shall accordingly restrict ourselves to the following
four dot eigenstates: the ground (or vacuum) state $ |G\rangle $;
the two optically active single-exciton states $ | X_R \rangle $
and $ | X_L \rangle $, with $z$ components of the overall angular
momentum $+1$ and $-1$, respectively; the lowest biexciton state $
| B \rangle $. The Coulomb interaction between the carriers,
enhanced by their 3D confinement, results in an energy
renormalization of the exciton transitions (the so called
biexciton binding energy) of the order of a few meV. These,
together with the polarization-dependent selection rules, allow a
selective addressing of the different transitions, already at the
ps timescale (see Fig.~\ref{lscheme}). In most self-assembled QDs,
the spin degeneracy of the bright exciton states is removed by the
anisotropic electron-hole exchange interaction, the actual
eigenstates thus being the linearly polarized $ | X_x \rangle $
and $ | X_y \rangle $. This geometric effect can however be
compensated (e.g., by an external magnetic field), thus recovering
the circularly polarized exciton basis \cite{bayer}. Since our
goal is the efficient emission of photon pairs, we consider the QD
to be embedded in a cavity having the fundamental mode with the
following essential property: {\it its frequency $\omega _C$ is
different from the transition energies between QD states, but as
close as possible to one half the energy difference between $ | G
\rangle $ and $ | B \rangle $} ($ \Delta _1 = - \Delta _2 $ in
Fig.~\ref{lscheme}). This {\it two-photon resonance} condition
seems promising for enhancing the emission of photon pairs with
respect to that produced by QD's without cavity
\cite{santori02,stevenson,ulrich03,ulrich}. The cavity mode also
presents a degeneracy with respect to the right and left
polarizations. Although we do not impose any constriction on the
number of photons inside the cavity, their number is limited to
only a few units with the adopted range of physical parameters.

In order to account for the open nature of the cavity-dot system, we simulate
its dynamics by means of a density-matrix description. In particular, the
evolution of the density operator $\rho$ is given by the following master
equation in the Lindblad form~\cite{scully} ($\hbar = 1$):
\begin{eqnarray}
\frac{d}{dt}  \rho & = & i \left[ \rho, H_S \right] + \sum
_{J=R,L} \frac{\kappa}{2} \left( 2 a_J \rho a^{\dagger}_J -
a^{\dagger}_J a_J \rho - \rho a^{\dagger}_J a_J  \right) +
\frac{\gamma}{2} \sum_{i=1}^4 \sum_{j=1}^4 \left( 2 \sigma_i \rho
\sigma^{\dagger}_j - \sigma^{\dagger}_i \sigma_j \rho - \rho
\sigma^{\dagger}_i \sigma_j \right)
\nonumber \\
& & \times (\delta_{i,j} + \delta_{|i-j|,2}) + \frac{P}{2}
\sum_{n=1}^{N_P} \theta [t-(n-1)T+t_P] \ \theta [(n-1)T-t]
\sum_{i=1}^4 \sum_{j=1}^4 \left( 2 \sigma^{\dagger}_i \rho
\sigma_j \right.
\nonumber \\
& & \left. - \sigma_i \sigma^{\dagger}_j\rho - \rho \sigma_i
\sigma^{\dagger}_j \right) (\delta_{i,j} + \delta_{|i-j|,2}),
\label{masterequation}
\end{eqnarray}
where $a_{J}$ ($a_{J}^\dagger$) is the annihilation (creation) operator for a
photon with polarization $J=R,L$, while $\sigma_{i=1,4} = | G \rangle \langle
X_R |, | G \rangle \langle X_L |,| X_L \rangle \langle B |, | X_R \rangle
\langle B |$ are the ladder operators.
The second and third terms on the right-hand side of the equation account for
the radiative relaxation of the cavity and of the dot, respectively; the last
one corresponds to its pulsed, incoherent pumping (i.e., the electron-hole
pairs being photogenerated in the wetting layer, before relaxing non-radiatively
in the dot).
The Hamiltonian of the coupled QD-MC
system is
\begin{eqnarray}
H_{S} & = & (\omega _C + \Delta _1 ) \left[ \mid X_R \rangle \langle X_R \mid +
\mid X_L \rangle \langle X_L \mid \right] + (2 \omega _C + \Delta _1 + \Delta _2)
\mid B \rangle \langle B \mid
\nonumber \\
& & + \sum _{J=R,L} \left[ \omega _C \left( a^\dagger_J a_J
+ 1/2 \right) + \sum _{i} q \left( \sigma _i a^\dagger _J + a_J \sigma ^\dagger _i
\right) \right].
\end{eqnarray}
In the following, we shall consider the case of degenerate excitons, while
their transition energies are assumed to be detuned with respect to the
cavity-mode frequency: $ E_{X_{R,L}} - E_G = \omega_C + \Delta_1 $ and $ E_B -
E_{X_{R,L}} = \omega_C + \Delta_2 $. As already mentioned, we will focus on the
two-photon resonance case, $ \Delta_1 = - \Delta_2 $. Besides, the light
emission through the leaky modes is considered inefficient as compared to that
from the MC ($ \gamma \ll \kappa $). Finally, the system is pumped by means of
rectangular pulses, of intensity $ P $, duration $t_P$, and repetition rate
$ 1/T $.

The correlation properties of the emitted radiation are described
by the first- and second-order coherence functions. Photon
correlations outside the cavity can be considered proportional to
those inside~\cite{walls,stace}. Therefore, the
polarization-resolved second-order coherence functions we shall
refer to in the following are:
\begin{equation}
g_{J,J'}^{(2)}(t,t+\tau) =
\frac{G_{J,J'}^{(2)}(t,t+\tau)}{\langle
a_{J}^\dagger(t)a_{J}(t)\rangle \langle
a_{J'}^\dagger(t+\tau)a_{J'}(t+\tau)\rangle} \, ,
\label{g2}
\end{equation}
where
\begin{equation}
G_{J,J'}^{(2)}(t,t+\tau) = \langle a_{J}^\dagger(t)
a_{J'}^\dagger(t+\tau) a_{J'}(t+\tau)a_{J}(t) \rangle \, ,
\label{G2}
\end{equation}
$J,J' = R,L$ being the photon polarizations. The dependence of the
$G^{(2)}_{J,J'} (t,t+\tau)$ on the delay $ \tau $ is derived,
given the system state $\rho (t)$, by means of the quantum
regression theorem~\cite{scully}. Further details on the method can
be found in Ref.~\cite{perea04}.

\section{ RESULTS}
\subsection{Second order coherence functions}
\begin{figure}
\includegraphics[clip,height=8cm,width=8.cm]{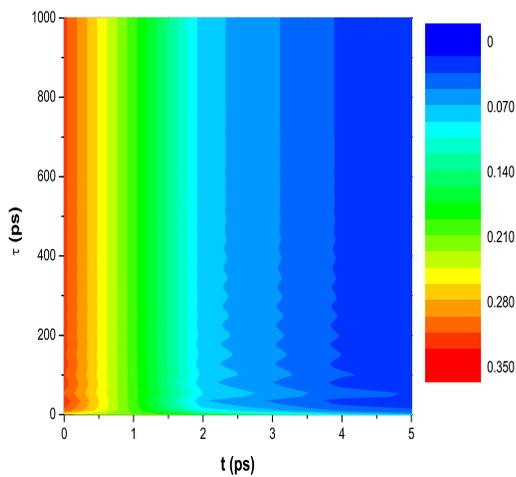}
\caption{(Color online) Second-order coherence function
$g_{L,L}^{(2)} ( t , \tau ) $ as a function of the two time
arguments (in ps). The time origin $ t = 0 $ is fixed at the end
of the first laser pulse. The values of the physical parameters
are: $ q = 0.1 $~meV, $ \kappa = 0.1 $~meV, $\Delta_1 = - \Delta_2
= 0.5 $~meV, $\gamma = 0.01 $~meV, $ P = 1$~meV, and $ t_P = 3
$~ps.} \label{g2ll}
\end{figure}

\begin{figure}
\includegraphics[clip,height=8cm,width=8.cm]{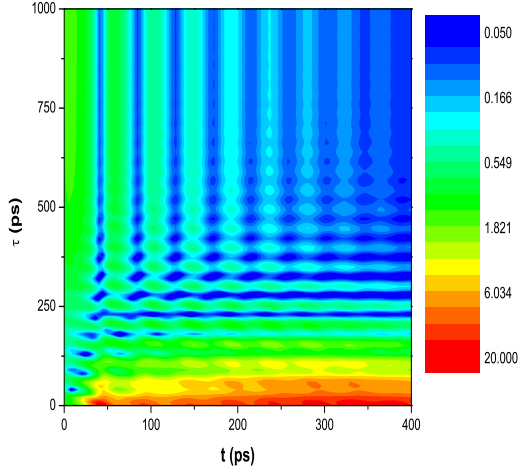}
\caption{(Color on line) Second-order coherence function $
g_{R,L}^{(2)} (t,\tau)$ as a function of the two time arguments
(in ps). The values of the physical parameters are the ones
reported in the caption of Fig.~\ref{g2ll}.} \label{g2rl}
\end{figure}

We start by illustrating the overall time-dependence of the second-order
coherence functions, for typical values of the relevant physical parameters
(specified in the figure captions), and a single excitation pulse ($N_P=1$).
Figures~\ref{g2ll} and \ref{g2rl} show $ g_{L,L}^{(2)} (t,\tau) $ and
$ g_{R,L}^{(2)} (t,\tau) $, respectively ($ g_{R,R}^{(2)} = g_{L,L}^{(2)} $
and $ g_{L,R}^{(2)} = g_{R,L}^{(2)} $ due to the system's symmetry with
respect to light and exciton polarizations). The emitted light
clearly exhibits non-classical features. The function $ g_{L,L}^{(2)} $ quite
generally shows a strongly sub-poissonian statistics [$ g^{(2)} (t,0) < 1 $],
and a clear anti-bunching behavior [$ g^{(2)} (t,\tau) > g^{(2)} (t,0) $]. In
fact, even though the present excitation regime is high enough for the system
to be multiply excited, the polarization properties of the biexciton states
and the optical selection rules suppress the probability for more than one
photon to be emitted with the same polarization. Besides, due to the pulsed
nature of the system's excitation, the memory of the first photon being
collected is never lost: correspondingly, the asymptotic value the $ g^{(2)}
(t,\tau) $ tends to for infinite delay is always well below 1, though larger
than $ g^{(2)} (t,0) $. The value of 1, which characterizes the
continuous-pumping case~\cite{perea05}, is recovered for increasing duration
$t_P$ of the pumping pulse (not shown here). As may be expected, the features
emerging from the analysis of the $ g_{R,L}^{(2)} (t,\tau) $ function are
quite different. In fact, the emissions of the $R$ and $L$ photons are
positively correlated, and a strongly super-poissonian statistics emerges. The
asymptotic values $g_{R,L}^{(2)} (t,\tau \rightarrow \infty)$, depend on the
initial time $t$, whereas the overall dependence on $\tau$ includes both
bunching and anti-bunching
features.

\begin{figure}
\includegraphics[clip,height=8cm,width=8.cm,angle=-90]{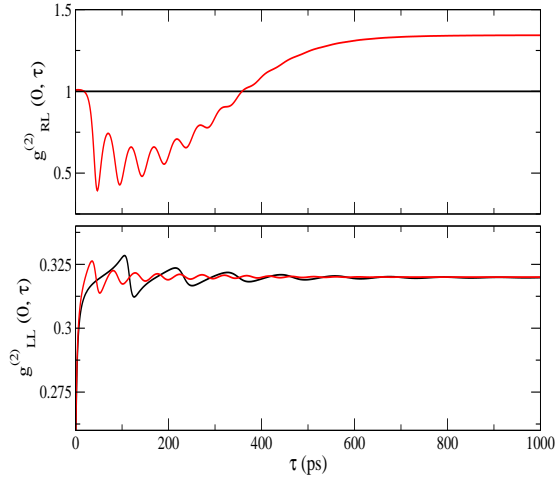}
\caption{Color online. Second order coherence functions
$g_{R,L}^{(2)}(0,\tau )$ (upper panel) and  $g_{L,L}^{(2)}(0,\tau
)$ (lower panel) as a function of the delay (in ps). The time
$t=0$ corresponds to the end of the squared pulse (see
Eq.~\ref{masterequation}). Red curves correspond to the {\it
two-photon resonance} situation $\Delta _1 = - \Delta _2 = 0.5
$~meV, while the black curves correspond to a non-resonant case
$\Delta _1 = \Delta _2 = 0.1 $~meV. The other parameters are
$q=0.1$~meV, $ \kappa =0.1$~meV, $ \gamma = 0.01 $~meV, $ P = 0.1
$~meV and $ t_P = 3 $~ps.} \label{todoenuno}
\end{figure}

In Fig.~\ref{todoenuno} we plot the functions $ g_{J,J'}^{(2)} (0,\tau) $, in
order to better appreciate some of the details, and to compare the two-photon
resonance case $\Delta _1=-\Delta_2=0.5$~meV with one where such condition is
not fulfilled ($\Delta _1=\Delta _2=0.1$~meV). In the two-photon resonance case,
$(J,J')=(R,L)$ exhibits the above-mentioned rich behavior (upper panel): it is
close to 1 (Poissonian statistics) at zero delay; it then oscillates with a
pseudo period of the about 50~ps, while remaining below its initial value
(photon bunching); finally, it reaches an asymptotic value of about $1.3$ (red
curve).

On the contrary, for $ \Delta_1
\neq - \Delta_2 $ (black curve), any correlation between the light emission in
the two polarizations is suppressed ($g^{(2)}_{R,L} (0,\tau )=1$). The case
$ ( J , J' ) = ( L , L ) $ is shown in the lower panel: here,
the two-photon resonance doesn't play
any role, because it doesn't apply to two photons with the same polarization.
The deep at zero delay demonstrates that the multiple excitation of each cavity
mode is completely negligible, whereas that of the dot-cavity system is not
(red curve): an $L$ excitation might still be transferred from the QD to the MC
after an $L$ photon has been emitted at $t=0$. In this case, the differences
with respect to the off-resonance case (black curve) is only due to the
difference between the values of $\Delta_{1,2}$.

\subsection{Two-photon coincidences}

The behaviors sofar discussed are not directly accessible in
experiments, for they occur on timescales which are shorter than
those presently achievable, e.g., within a Hanbury-Brown-Twiss
setup~\cite{scully}. In fact, the single photon detectors
currently used in this kind of experiments have detection times
$t_{det}$ of the order of hundreds of ps, whereas the typical
timescales $G_{J,J'}^{(2)}(t,t+\tau) $ evolves on are of tens of
ps. Therefore, correlation functions have to be integrated on time
intervals of the order of $t_{det}$, or larger, and normalized by
analogous integrals of the populations $ \langle a_J^{\dagger} (t)
a_J (t) \rangle $.

\begin{figure}
\includegraphics[clip,height=8cm,width=8.cm,angle=-90]{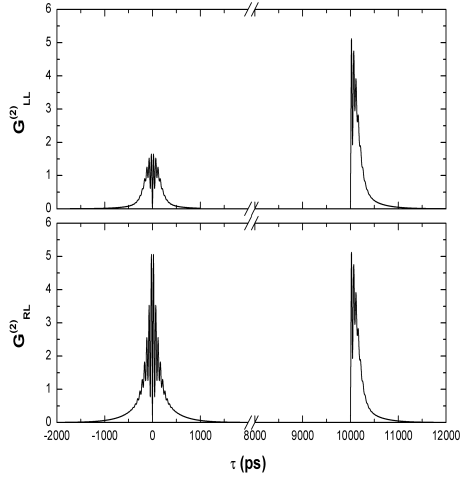}
\caption{Non-normalized coincidences as given by
$G_{J,J'}^{(2)}(0,\tau)$ as a function of the delay (in ps) for a
start time $t$ just at the end of a pulse. The two pulses are
separated by $ T = 10 $~ns. The values of the physical parameters
are: $ q = 0.1 $~meV, $ \kappa = 0.1 $~meV, $ \Delta _1 = - \Delta
_2 = 0.5 $~meV, $ \gamma = 0.01 $~meV, $ P = 1 $~meV, and $ t_P =
3 $~ps.} \label{2pulses}
\end{figure}

Experimentally, the second-order correlation functions can be normalized
by repeatedly exciting the dot by means of identical pulses, separated
by time intervals $T$ of the order of $10$~ns; such a value is supposed to
be much larger than any correlation time in the system. Therefore, when
the start and the stop detections correspond to different pulses ($\tau >
T$), no correlation at all is expected. By normalizing
the second order correlation $G_{J,J'}^{(2)}
(t,t+\tau) $ with such \emph{coincidences at large delay}, the value of 1
is obtained in the case of a pulsed laser~\cite{santori}.
As a first step in the investigation of this aspect, we perform the
calculation of $G_{J,J'}^{(2)}(t,t+\tau)$, for $ \tau > T $.
This function is shown in Fig.~\ref{2pulses} for $t=0$, $T=10$~ns,
and all the other parameters as in Figs.~\ref{g2ll} and
\ref{g2rl}; similar results are obtained for any other time $t$
after the first pulse.

\begin{figure}
\includegraphics[clip,height=8cm,width=8.cm,angle=-90]{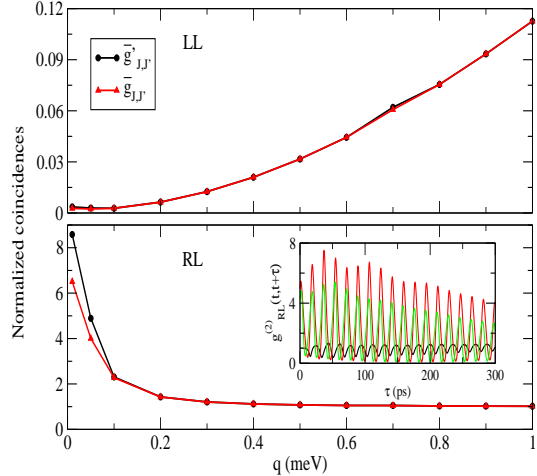}
\caption{(Color on line) Normalized coincidences as given by
$\overline{g}_{J,J'}$ (red line) or the approximation
$\overline{g}_{J,J'}'$ (black line) as a function of $q$. The
values of the parameters are: $\kappa =0.1 $~meV, $\Delta _1
=-\Delta _2 =0.5$~meV, $\gamma =0.01$~meV, $P=1$~meV, $t_P =3$~ps
and $T=10$~ns. The inset shows $g_{R,L}^{(2)}(t,t+\tau)$ as a
function of $\tau$ for $q=0.7 $~meV and $t=0$, $50$ and $100$~ps.}
\label{inter_G2}
\end{figure}

In the following we use the above results in order to quantify the
two-photon correlations. The ideal
normalization of $ G_{J,J'}^{(2)} $ is given by the time integral of the
uncorrelated second-order correlation function, which is given by
$ \langle n_J (t) \rangle \, \langle n_{J'} (t+\tau ) \rangle $:
\begin{equation}
\overline{g}_{J,J'} = \frac{\int_{0}^{T} dt \int_{0}^{T} d\tau
G_{J,J'}^{(2)}(t,t+\tau) } { \int_{0}^{T} dt    \langle
a_{J}^\dagger(t) a_{J}(t) \rangle \int_{0}^{T} dt \langle
a_{J'}^\dagger(t) a_{J'}(t)\rangle } . \label{G2nor3}
\end{equation}
In practice, if the dot is periodically excited by sequences of identical
laser pulses, and if the time interval $T$ separating two consecutive such
pulses is larger than the system's memory, then
$ G_{J,J'}^{(2)}(t,t+\tau) \simeq \langle n_J (t) \rangle \, \langle n_{J'}
(t+\tau ) \rangle $ for $ \tau > T $.
Correspondingly, $ \overline{g}_{J,J'} $ can be approximated by
\begin{equation}
\overline{g}_{J,J'}' = \frac{\int_{0}^{T} dt \int_{0}^{T} d\tau \;
G_{J,J'}^{(2)}(t,t+\tau) } { \int_{t}^{T} dt \int_{T-t}^{2T-t}
d\tau \; G_{J,J'}^{(2)} (t,t+\tau)} . \label{G2nor2}
\end{equation}
In other words, two-photon correlation is approximated by the
ratio between the areas of the first and the second peaks in
Fig.~\ref{2pulses}. Our simulations of the system's evolution
under the effect of two identical squared pulses ($N_P =2$) aim at
understanding to which extent such an approximation actually
holds. Being the overall integration time $2T$ orders of magnitude
larger than the characteristic timescales of the dot-cavity
dynamics and, thus, of the time-step of the numerical
calculations, such simulations are extremely time-consuming. In
Fig.~\ref{inter_G2} we compare the values of $\overline{g}_{J,J'}$
and $\overline{g}_{J,J'}'$, for different system parameters, while
the time delay between consecutive laser pulses is kept constant
($T=10$~ns).

Our results show that, for small values of the dot-cavity coupling
$q$, the approximation of $\overline{g}_{J,J'}'$ with
$\overline{g}_{J,J'}$ is not completely adequate: although the
occupations of all the excited states in the system decay to
negligible values on timescales of the order of $T$, the coherence
function $G_{J,J'}^{(2)}(t,t+\tau)$ still cannot be factorized,
for its decay with $\tau$ is slower than that of the density
matrix with $t$.

Aside from the normalization issue, a strong dependence of the
second-order correlations on the dot-cavity coupling constant
emerges. The value of $ \overline{g}_{L,L} = \overline{g}_{R,R} $
monotonically increases with $q$, while it remains well below 1 in
all the range of considered parameters. This confirms the system's
general tendency to emit not more than one photon of each
polarization. In the limit of weak dot-cavity coupling, and thus
of slow excitation transfer from the QD to the MC, the probability
for the system to be further excited after the first photon
emission has taken place is practically suppressed. In order to
identify the weak and strong coupling regimes, one can make the
following simple argument: the QD requires a time $ \sim 1 / P $
to be excited, and $ \sim 1 / q $ to transfer its excitation to the
MC. Therefore, if $ 1 / q \ll t_P, 1 / P \sim 4$~ps, there is no time
for the dot to relax by emitting a photon (in the cavity) and being
subsequently re-excited by the laser pulse. In this respect, the
region where $ q < 0.2 $~meV can be identified with the weak-coupling
regime.

On the other hand, $ \overline{g}_{R,L} $ decreases with
increasing $q$. In particular, a strong correlation is observed at
the weak coupling regime, whereas the probability of an $L$
photon being emitted becomes nearly independent from the previous
observation of $R$ ones ($ \overline{g}_{R,L} \sim 1$) for $q
\gtrsim 0.2$~meV.
One could be tempted to assign some coherence properties to the
regime $ \overline{g}_{R,L} \sim 1$, being this the
experimental value characteristic of a pulsed laser~\cite{santori}.
However, as shown in the inset of Fig.~\ref{inter_G2},
this is not the case for $g_{R,L}^{(2)}(t,t+\tau)$. In fact, the second-order
coherence function shows strong and fast oscillations, indicating no
coherence at all. Therefore,
$\overline{g}_{R,L}=1$ cannot be identified with coherence.

\section{CONCLUSIONS}
We have studied the photon emission of a QD, embedded in a MC and
incoherently excited by pulsed pumping. We have shown that in the
two-photon resonance condition, strongly positive correlations
between $R$ and $L$ radiation can be achieved, while keeping
negligible the probability of emitting multiple, equally polarized
photons. Under proper excitation conditions, thus, the QD behaves
as an efficient emitter of counter-polarized photon pairs. The
detailed analysis of the second-order coherence functions
$g_{J,J'}^{(2)}$ shows a rich behavior, including strong
oscillations on a $10$~ps timescale; the asymptotic values ($\tau
\rightarrow \infty$) differ from 1, the value expected in the
continuous-pumping case. Time-averaged correlations $
\overline{g}_{J,J'} $ have also been computed, in order to allow a
more direct comparison with experimental results. The
above-mentioned features are clearly reflected also in their
values, specially for small values of the dot-cavity coupling
constant $q$.

\section{ACKNOWLEDGEMENTS}
This work has been partly supported by the Spanish MCyT under
contract No. MAT2002-00139, CAM under Contract No. 07N/0042/2002,
and the European Union within the Research Training Network
COLLECT.


\bibliography{pulses}

\begin{thebibliography}{24}
\expandafter\ifx\csname natexlab\endcsname\relax\def\natexlab#1{#1}\fi
\expandafter\ifx\csname bibnamefont\endcsname\relax
  \def\bibnamefont#1{#1}\fi
\expandafter\ifx\csname bibfnamefont\endcsname\relax
  \def\bibfnamefont#1{#1}\fi
\expandafter\ifx\csname citenamefont\endcsname\relax
  \def\citenamefont#1{#1}\fi
\expandafter\ifx\csname url\endcsname\relax
  \def\url#1{\texttt{#1}}\fi
\expandafter\ifx\csname urlprefix\endcsname\relax\def\urlprefix{URL }\fi
\providecommand{\bibinfo}[2]{#2}
\providecommand{\eprint}[2][]{\url{#2}}

\bibitem[{\citenamefont{Moreau et~al.}(2001{\natexlab{a}})\citenamefont{Moreau,
  Robert, Gerard, Abram, Manin, and Thierry-Mieg}}]{moreau}
\bibinfo{author}{\bibfnamefont{E.}~\bibnamefont{Moreau}},
  \bibinfo{author}{\bibfnamefont{I.}~\bibnamefont{Robert}},
  \bibinfo{author}{\bibfnamefont{J.~M.} \bibnamefont{Gerard}},
  \bibinfo{author}{\bibfnamefont{I.}~\bibnamefont{Abram}},
  \bibinfo{author}{\bibfnamefont{L.}~\bibnamefont{Manin}}, \bibnamefont{and}
  \bibinfo{author}{\bibfnamefont{V.}~\bibnamefont{Thierry-Mieg}},
  \bibinfo{journal}{App. Phys. Lett.} \textbf{\bibinfo{volume}{79}},
  \bibinfo{pages}{2865} (\bibinfo{year}{2001}{\natexlab{a}}).

\bibitem[{\citenamefont{Santori
  et~al.}(2002{\natexlab{a}})\citenamefont{Santori, Fattal, Vuckovic, Solomon,
  and Yamamoto}}]{santori}
\bibinfo{author}{\bibfnamefont{C.}~\bibnamefont{Santori}},
  \bibinfo{author}{\bibfnamefont{D.}~\bibnamefont{Fattal}},
  \bibinfo{author}{\bibfnamefont{J.}~\bibnamefont{Vuckovic}},
  \bibinfo{author}{\bibfnamefont{G.}~\bibnamefont{Solomon}}, \bibnamefont{and}
  \bibinfo{author}{\bibfnamefont{Y.}~\bibnamefont{Yamamoto}},
  \bibinfo{journal}{Nature} \textbf{\bibinfo{volume}{419}},
  \bibinfo{pages}{594} (\bibinfo{year}{2002}{\natexlab{a}}).

\bibitem[{\citenamefont{Nielsen and Chuang}(2000)}]{nielsen}
\bibinfo{author}{\bibfnamefont{M.}~\bibnamefont{Nielsen}} \bibnamefont{and}
  \bibinfo{author}{\bibfnamefont{I.}~\bibnamefont{Chuang}}, in
  \emph{\bibinfo{booktitle}{Quantum Computation and Quantum Information}}
  (\bibinfo{publisher}{Cambridge University Press, Cambridge},
  \bibinfo{year}{2000}).

\bibitem[{\citenamefont{Solomon et~al.}(2001)\citenamefont{Solomon, Pelton, and
  Yamamoto}}]{solomon}
\bibinfo{author}{\bibfnamefont{G.~S.} \bibnamefont{Solomon}},
  \bibinfo{author}{\bibfnamefont{M.}~\bibnamefont{Pelton}}, \bibnamefont{and}
  \bibinfo{author}{\bibfnamefont{Y.}~\bibnamefont{Yamamoto}},
  \bibinfo{journal}{Phys. Rev. Lett.} \textbf{\bibinfo{volume}{86}},
  \bibinfo{pages}{3903} (\bibinfo{year}{2001}).

\bibitem[{\citenamefont{Gerard et~al.}(1998)\citenamefont{Gerard, Sermage,
  Gayral, Legrand, Costard, and Thierry-Mieg}}]{gerard}
\bibinfo{author}{\bibfnamefont{J.~M.} \bibnamefont{Gerard}},
  \bibinfo{author}{\bibfnamefont{B.}~\bibnamefont{Sermage}},
  \bibinfo{author}{\bibfnamefont{B.}~\bibnamefont{Gayral}},
  \bibinfo{author}{\bibfnamefont{B.}~\bibnamefont{Legrand}},
  \bibinfo{author}{\bibfnamefont{E.}~\bibnamefont{Costard}}, \bibnamefont{and}
  \bibinfo{author}{\bibfnamefont{V.}~\bibnamefont{Thierry-Mieg}},
  \bibinfo{journal}{Phys. Rev. Lett.} \textbf{\bibinfo{volume}{81}},
  \bibinfo{pages}{1110} (\bibinfo{year}{1998}).

\bibitem[{\citenamefont{Unitt et~al.}(2005)\citenamefont{Unitt, Bennett,
  Atkinson, Ritchie, and Shields}}]{unitt}
\bibinfo{author}{\bibfnamefont{D.~C.} \bibnamefont{Unitt}},
  \bibinfo{author}{\bibfnamefont{A.~J.} \bibnamefont{Bennett}},
  \bibinfo{author}{\bibfnamefont{P.}~\bibnamefont{Atkinson}},
  \bibinfo{author}{\bibfnamefont{D.~A.} \bibnamefont{Ritchie}},
  \bibnamefont{and} \bibinfo{author}{\bibfnamefont{A.~J.}
  \bibnamefont{Shields}}, \bibinfo{journal}{Phys. Rev. B}
  \textbf{\bibinfo{volume}{72}}, \bibinfo{pages}{033318}
  (\bibinfo{year}{2005}).

\bibitem[{\citenamefont{Varoutsis et~al.}(2005)\citenamefont{Varoutsis,
  Laurent, Kramper, Lemaitre, Sagnes, Robert-Philip, and Abram}}]{varoutsis}
\bibinfo{author}{\bibfnamefont{S.}~\bibnamefont{Varoutsis}},
  \bibinfo{author}{\bibfnamefont{S.}~\bibnamefont{Laurent}},
  \bibinfo{author}{\bibfnamefont{P.}~\bibnamefont{Kramper}},
  \bibinfo{author}{\bibfnamefont{A.}~\bibnamefont{Lemaitre}},
  \bibinfo{author}{\bibfnamefont{I.}~\bibnamefont{Sagnes}},
  \bibinfo{author}{\bibfnamefont{I.}~\bibnamefont{Robert-Philip}},
  \bibnamefont{and} \bibinfo{author}{\bibfnamefont{I.}~\bibnamefont{Abram}},
  \bibinfo{journal}{Phys. Rev. B} \textbf{\bibinfo{volume}{72}},
  \bibinfo{pages}{041303(R)} (\bibinfo{year}{2005}).

\bibitem[{\citenamefont{Moreau et~al.}(2001{\natexlab{b}})\citenamefont{Moreau,
  Robert, Manin, Thierry-Mieg, Gerard, and Abram}}]{moreau01}
\bibinfo{author}{\bibfnamefont{E.}~\bibnamefont{Moreau}},
  \bibinfo{author}{\bibfnamefont{I.}~\bibnamefont{Robert}},
  \bibinfo{author}{\bibfnamefont{L.}~\bibnamefont{Manin}},
  \bibinfo{author}{\bibfnamefont{V.}~\bibnamefont{Thierry-Mieg}},
  \bibinfo{author}{\bibfnamefont{J.~M.} \bibnamefont{Gerard}},
  \bibnamefont{and} \bibinfo{author}{\bibfnamefont{I.}~\bibnamefont{Abram}},
  \bibinfo{journal}{Phys. Rev. Lett.} \textbf{\bibinfo{volume}{87}},
  \bibinfo{pages}{183601} (\bibinfo{year}{2001}{\natexlab{b}}).

\bibitem[{\citenamefont{Santori et~al.}(2001)\citenamefont{Santori, Pelton,
  Solomon, Dale, and Yamamoto}}]{santori01}
\bibinfo{author}{\bibfnamefont{C.}~\bibnamefont{Santori}},
  \bibinfo{author}{\bibfnamefont{M.}~\bibnamefont{Pelton}},
  \bibinfo{author}{\bibfnamefont{G.}~\bibnamefont{Solomon}},
  \bibinfo{author}{\bibfnamefont{Y.}~\bibnamefont{Dale}}, \bibnamefont{and}
  \bibinfo{author}{\bibfnamefont{Y.}~\bibnamefont{Yamamoto}},
  \bibinfo{journal}{Phys. Rev. Lett.} \textbf{\bibinfo{volume}{86}},
  \bibinfo{pages}{1502} (\bibinfo{year}{2001}).

\bibitem[{\citenamefont{Zwiller et~al.}(2002)\citenamefont{Zwiller, Jonsson,
  Blom, Jeppesen, Pistol, Samuelson, Katznelson, Kotelnikov, Evtikhiev, and
  Bjork}}]{zwiller02}
\bibinfo{author}{\bibfnamefont{V.}~\bibnamefont{Zwiller}},
  \bibinfo{author}{\bibfnamefont{P.}~\bibnamefont{Jonsson}},
  \bibinfo{author}{\bibfnamefont{H.}~\bibnamefont{Blom}},
  \bibinfo{author}{\bibfnamefont{S.}~\bibnamefont{Jeppesen}},
  \bibinfo{author}{\bibfnamefont{M.~E.} \bibnamefont{Pistol}},
  \bibinfo{author}{\bibfnamefont{L.}~\bibnamefont{Samuelson}},
  \bibinfo{author}{\bibfnamefont{A.~A.} \bibnamefont{Katznelson}},
  \bibinfo{author}{\bibfnamefont{E.~Y.} \bibnamefont{Kotelnikov}},
  \bibinfo{author}{\bibfnamefont{V.}~\bibnamefont{Evtikhiev}},
  \bibnamefont{and} \bibinfo{author}{\bibfnamefont{G.}~\bibnamefont{Bjork}},
  \bibinfo{journal}{Phys. Rev. A} \textbf{\bibinfo{volume}{66}},
  \bibinfo{pages}{53814} (\bibinfo{year}{2002}).

\bibitem[{\citenamefont{Kiraz et~al.}(2002)\citenamefont{Kiraz, Falth, Becher,
  Gayral, Schoenfeld, Petroff, Zhang, Hu, and Imamoglu}}]{kiraz02}
\bibinfo{author}{\bibfnamefont{A.}~\bibnamefont{Kiraz}},
  \bibinfo{author}{\bibfnamefont{S.}~\bibnamefont{Falth}},
  \bibinfo{author}{\bibfnamefont{C.}~\bibnamefont{Becher}},
  \bibinfo{author}{\bibfnamefont{B.}~\bibnamefont{Gayral}},
  \bibinfo{author}{\bibfnamefont{W.~V.} \bibnamefont{Schoenfeld}},
  \bibinfo{author}{\bibfnamefont{P.~M.} \bibnamefont{Petroff}},
  \bibinfo{author}{\bibfnamefont{L.}~\bibnamefont{Zhang}},
  \bibinfo{author}{\bibfnamefont{E.}~\bibnamefont{Hu}}, \bibnamefont{and}
  \bibinfo{author}{\bibfnamefont{A.}~\bibnamefont{Imamoglu}},
  \bibinfo{journal}{Phys. Rev. B} \textbf{\bibinfo{volume}{65}},
  \bibinfo{pages}{161303(R)} (\bibinfo{year}{2002}).

\bibitem[{\citenamefont{Michler et~al.}(2000)\citenamefont{Michler, Kiraz,
  Becher, Schoenfeld, Petroff, Zhang, Hu, and Imamoglu}}]{michler}
\bibinfo{author}{\bibfnamefont{P.}~\bibnamefont{Michler}},
  \bibinfo{author}{\bibfnamefont{A.}~\bibnamefont{Kiraz}},
  \bibinfo{author}{\bibfnamefont{C.}~\bibnamefont{Becher}},
  \bibinfo{author}{\bibfnamefont{W.}~\bibnamefont{Schoenfeld}},
  \bibinfo{author}{\bibfnamefont{P.}~\bibnamefont{Petroff}},
  \bibinfo{author}{\bibfnamefont{L.}~\bibnamefont{Zhang}},
  \bibinfo{author}{\bibfnamefont{E.}~\bibnamefont{Hu}}, \bibnamefont{and}
  \bibinfo{author}{\bibfnamefont{A.}~\bibnamefont{Imamoglu}},
  \bibinfo{journal}{Science} \textbf{\bibinfo{volume}{290}},
  \bibinfo{pages}{2282} (\bibinfo{year}{2000}).

\bibitem[{\citenamefont{Pelton et~al.}(2002)\citenamefont{Pelton, Santori,
  Vuckovic, Zhang, Solomon, Plant, and Yamamoto}}]{pelton}
\bibinfo{author}{\bibfnamefont{M.}~\bibnamefont{Pelton}},
  \bibinfo{author}{\bibfnamefont{C.}~\bibnamefont{Santori}},
  \bibinfo{author}{\bibfnamefont{J.}~\bibnamefont{Vuckovic}},
  \bibinfo{author}{\bibfnamefont{B.}~\bibnamefont{Zhang}},
  \bibinfo{author}{\bibfnamefont{G.}~\bibnamefont{Solomon}},
  \bibinfo{author}{\bibfnamefont{J.}~\bibnamefont{Plant}}, \bibnamefont{and}
  \bibinfo{author}{\bibfnamefont{Y.}~\bibnamefont{Yamamoto}},
  \bibinfo{journal}{Phys. Rev. Lett.} \textbf{\bibinfo{volume}{89}},
  \bibinfo{pages}{233602} (\bibinfo{year}{2002}).

\bibitem[{\citenamefont{Fattal et~al.}(2004)\citenamefont{Fattal, Inoue,
  Vuckovic, Santori, Solomon, and Yamamoto}}]{fattal04}
\bibinfo{author}{\bibfnamefont{D.}~\bibnamefont{Fattal}},
  \bibinfo{author}{\bibfnamefont{K.}~\bibnamefont{Inoue}},
  \bibinfo{author}{\bibfnamefont{J.}~\bibnamefont{Vuckovic}},
  \bibinfo{author}{\bibfnamefont{C.}~\bibnamefont{Santori}},
  \bibinfo{author}{\bibfnamefont{G.~S.} \bibnamefont{Solomon}},
  \bibnamefont{and} \bibinfo{author}{\bibfnamefont{Y.}~\bibnamefont{Yamamoto}},
  \bibinfo{journal}{Phys. Rev. Lett.} \textbf{\bibinfo{volume}{92}},
  \bibinfo{pages}{37903} (\bibinfo{year}{2004}).

\bibitem[{\citenamefont{Bayer et~al.}(2002)\citenamefont{Bayer, Ortner, Stern,
  Kuther, Gorbunov, Forchel, Hawrylak, Fafard, Hinzer, Reinecke
  et~al.}}]{bayer}
\bibinfo{author}{\bibfnamefont{M.}~\bibnamefont{Bayer}},
  \bibinfo{author}{\bibfnamefont{G.}~\bibnamefont{Ortner}},
  \bibinfo{author}{\bibfnamefont{O.}~\bibnamefont{Stern}},
  \bibinfo{author}{\bibfnamefont{A.}~\bibnamefont{Kuther}},
  \bibinfo{author}{\bibfnamefont{A.}~\bibnamefont{Gorbunov}},
  \bibinfo{author}{\bibfnamefont{A.}~\bibnamefont{Forchel}},
  \bibinfo{author}{\bibfnamefont{P.}~\bibnamefont{Hawrylak}},
  \bibinfo{author}{\bibfnamefont{S.}~\bibnamefont{Fafard}},
  \bibinfo{author}{\bibfnamefont{K.}~\bibnamefont{Hinzer}},
  \bibinfo{author}{\bibfnamefont{T.}~\bibnamefont{Reinecke}},
  \bibnamefont{et~al.}, \bibinfo{journal}{Phys. Rev. B}
  \textbf{\bibinfo{volume}{65}}, \bibinfo{pages}{195315}
  (\bibinfo{year}{2002}).

\bibitem[{\citenamefont{Santori
  et~al.}(2002{\natexlab{b}})\citenamefont{Santori, Fattal, Pelton, Solomon,
  and Yamamoto}}]{santori02}
\bibinfo{author}{\bibfnamefont{C.}~\bibnamefont{Santori}},
  \bibinfo{author}{\bibfnamefont{D.}~\bibnamefont{Fattal}},
  \bibinfo{author}{\bibfnamefont{M.}~\bibnamefont{Pelton}},
  \bibinfo{author}{\bibfnamefont{G.~S.} \bibnamefont{Solomon}},
  \bibnamefont{and} \bibinfo{author}{\bibfnamefont{Y.}~\bibnamefont{Yamamoto}},
  \bibinfo{journal}{Phys. Rev. B} \textbf{\bibinfo{volume}{66}},
  \bibinfo{pages}{45308} (\bibinfo{year}{2002}{\natexlab{b}}).

\bibitem[{\citenamefont{Stevenson et~al.}(2002)\citenamefont{Stevenson,
  Thompson, Shields, Farrer, Kardynal, Ritchie, and Pepper}}]{stevenson}
\bibinfo{author}{\bibfnamefont{R.~M.} \bibnamefont{Stevenson}},
  \bibinfo{author}{\bibfnamefont{R.~M.} \bibnamefont{Thompson}},
  \bibinfo{author}{\bibfnamefont{A.~J.} \bibnamefont{Shields}},
  \bibinfo{author}{\bibfnamefont{I.}~\bibnamefont{Farrer}},
  \bibinfo{author}{\bibfnamefont{B.~E.} \bibnamefont{Kardynal}},
  \bibinfo{author}{\bibfnamefont{D.~A.} \bibnamefont{Ritchie}},
  \bibnamefont{and} \bibinfo{author}{\bibfnamefont{M.}~\bibnamefont{Pepper}},
  \bibinfo{journal}{Phys. Rev. B} \textbf{\bibinfo{volume}{66}},
  \bibinfo{pages}{081302(R)} (\bibinfo{year}{2002}).

\bibitem[{\citenamefont{Ulrich et~al.}(2003)\citenamefont{Ulrich, Strauf,
  Michler, Bacher, and Forchel}}]{ulrich03}
\bibinfo{author}{\bibfnamefont{S.}~\bibnamefont{Ulrich}},
  \bibinfo{author}{\bibfnamefont{S.}~\bibnamefont{Strauf}},
  \bibinfo{author}{\bibfnamefont{P.}~\bibnamefont{Michler}},
  \bibinfo{author}{\bibfnamefont{G.}~\bibnamefont{Bacher}}, \bibnamefont{and}
  \bibinfo{author}{\bibfnamefont{A.}~\bibnamefont{Forchel}},
  \bibinfo{journal}{Appl. Phys. Lett.} \textbf{\bibinfo{volume}{83}},
  \bibinfo{pages}{1848} (\bibinfo{year}{2003}).

\bibitem[{\citenamefont{Ulrich et~al.}(2005)\citenamefont{Ulrich, Benyoucef,
  Michler, Baer, Gartner, adn M.~Schwab, H.Kurze, Bayer, Fafard, Wasilewski
  et~al.}}]{ulrich}
\bibinfo{author}{\bibfnamefont{S.}~\bibnamefont{Ulrich}},
  \bibinfo{author}{\bibfnamefont{M.}~\bibnamefont{Benyoucef}},
  \bibinfo{author}{\bibfnamefont{P.}~\bibnamefont{Michler}},
  \bibinfo{author}{\bibfnamefont{N.}~\bibnamefont{Baer}},
  \bibinfo{author}{\bibfnamefont{P.}~\bibnamefont{Gartner}},
  \bibinfo{author}{\bibfnamefont{F.~J.} \bibnamefont{adn M.~Schwab}},
  \bibinfo{author}{\bibnamefont{H.Kurze}},
  \bibinfo{author}{\bibfnamefont{M.}~\bibnamefont{Bayer}},
  \bibinfo{author}{\bibfnamefont{S.}~\bibnamefont{Fafard}},
  \bibinfo{author}{\bibfnamefont{Z.}~\bibnamefont{Wasilewski}},
  \bibnamefont{et~al.}, \bibinfo{journal}{Phys. Rev. B}
  \textbf{\bibinfo{volume}{71}}, \bibinfo{pages}{235328}
  (\bibinfo{year}{2005}).

\bibitem[{\citenamefont{Scully and Zubairy}(1997)}]{scully}
\bibinfo{author}{\bibfnamefont{M.}~\bibnamefont{Scully}} \bibnamefont{and}
  \bibinfo{author}{\bibfnamefont{M.}~\bibnamefont{Zubairy}}, in
  \emph{\bibinfo{booktitle}{Quantum optics}} (\bibinfo{publisher}{Cambridge
  University Press, Cambridge}, \bibinfo{year}{1997}).

\bibitem[{\citenamefont{Walls and Milburn}(1994)}]{walls}
\bibinfo{author}{\bibfnamefont{D.}~\bibnamefont{Walls}} \bibnamefont{and}
  \bibinfo{author}{\bibfnamefont{G.}~\bibnamefont{Milburn}}, in
  \emph{\bibinfo{booktitle}{Quantum optics}}
  (\bibinfo{publisher}{Springer-Verlag, Berlin}, \bibinfo{year}{1994}).

\bibitem[{\citenamefont{Stace et~al.}(2003)\citenamefont{Stace, Milburn, and
  Barnes}}]{stace}
\bibinfo{author}{\bibfnamefont{T.~M.} \bibnamefont{Stace}},
  \bibinfo{author}{\bibfnamefont{G.~J.} \bibnamefont{Milburn}},
  \bibnamefont{and} \bibinfo{author}{\bibfnamefont{C.~H.~W.}
  \bibnamefont{Barnes}}, \bibinfo{journal}{Phys. Rev. B}
  \textbf{\bibinfo{volume}{67}}, \bibinfo{pages}{085317}
  (\bibinfo{year}{2003}).

\bibitem[{\citenamefont{Perea et~al.}(2004)\citenamefont{Perea, Porras, and
  Tejedor}}]{perea04}
\bibinfo{author}{\bibfnamefont{J.~I.} \bibnamefont{Perea}},
  \bibinfo{author}{\bibfnamefont{D.}~\bibnamefont{Porras}}, \bibnamefont{and}
  \bibinfo{author}{\bibfnamefont{C.}~\bibnamefont{Tejedor}},
  \bibinfo{journal}{Phys. Rev. B} \textbf{\bibinfo{volume}{70}},
  \bibinfo{pages}{115304} (\bibinfo{year}{2004}).

\bibitem[{\citenamefont{Perea and Tejedor}(2005)}]{perea05}
\bibinfo{author}{\bibfnamefont{J.~I.} \bibnamefont{Perea}} \bibnamefont{and}
  \bibinfo{author}{\bibfnamefont{C.}~\bibnamefont{Tejedor}},
  \bibinfo{journal}{Phys. Rev. B} \textbf{\bibinfo{volume}{72}},
  \bibinfo{pages}{35303} (\bibinfo{year}{2005}).

\end{thebibliography}


\end{document}